\documentclass{aastex}
\usepackage{emulateapj5}
\submitted{Draft Version October 8, 2003}

\shorttitle{Lensing Cluster of Galaxies CL0024+17}
\shortauthors{Ota et al.}

\begin{document}

\title{{\it Chandra} Analysis and Mass Estimation of 
the Lensing Cluster of Galaxies CL0024+17}

\author{Naomi Ota\altaffilmark{1,2,3}, 
Etienne Pointecouteau\altaffilmark{4}, 
Makoto Hattori\altaffilmark{5}, 
and
Kazuhisa Mitsuda\altaffilmark{3}}

\altaffiltext{1}{Department of Physics, Tokyo Metropolitan University, 
1-1 Minami-osawa, Hachioji, Tokyo 192-0397, Japan}
\altaffiltext{2}{naomi@phys.metro-u.ac.jp}
\altaffiltext{3}{Institute of Space and Astronautical Science, 
Sagamihara, Kanagawa 229-8510, Japan}
\altaffiltext{4}{CEA/Saclay, L'Orme des Merisiers, 
91191 Gif sur Yvette, France}
\altaffiltext{5}{Tohoku University, Aoba Aramaki, Sendai 980-8578, Japan}

\begin{abstract} 
We present a detailed analysis of {\it Chandra} X-ray observations of
the lensing cluster of galaxies CL0024+17 at $z=0.395$. We found that
the radial temperature profile is consistent with being isothermal out
to $\sim 600$ kpc and that the average X-ray temperature is
$4.47^{+0.83}_{-0.54}$ keV.  The X-ray surface brightness profile is
represented by the sum of extended emission centered at the central
bright elliptical galaxy with a small core of 50 kpc and more extended
emission which can be well described by a spherical $\beta$-model with
a core radius of about 210 kpc.  Assuming the X-ray emitting gas to be
in hydrostatic equilibrium, we estimated the X-ray mass within the arc
radius and found it is significantly smaller than the strong lensing
mass by a factor of about 2--3. We detected a strong redshifted iron K
line in the X-ray spectrum from the cluster for the first time and
find the metal abundance to be $0.76^{+0.37}_{-0.31}$ solar.
\end{abstract}


\section{Introduction}
CL0024+17 is one of the most extensively studied distant clusters of
galaxies, located at $z=0.395$. The cluster is known to exhibit
spectacular multiple arc images of a background galaxy produced by the
gravitational lens effect. This makes the cluster a unique target for
studying the distribution of not only luminous matter but also dark
matter.

The lensed arc system was initially noted by \cite{Koo_1988} and the
first spectroscopic observation placed limits on the arc redshift,
$1<z_{\rm arc}<2$ \citep{Mellier_etal_1991}.  From the deep HST
imaging, four arc images of the blue source galaxy were identified
around the tangential critical curve and further was a radial arc of
the same source \citep{Smail_etal_1996, Colley_etal_1996}.  In
addition, CL0024+17 is the first cluster where a coherent weak shear
signal was detected \citep{Bonnet_etal_1994}.

The remarkably well-resolved lens images allow detailed modeling of
the dark matter distribution in the cluster center and several authors
have attempted the mass reconstruction \citep{Kassiola_etal_1992,
Smail_etal_1997, Tyson_etal_1998, Broadhurst_etal_2000}.
\cite{Tyson_etal_1998} has built a very detailed mass map comprising
512 free parameters. They found that, excluding the mass associated
with individual galaxies, the dark mass distribution is smooth and has
little asymmetry, and the central profile has a soft core of $35
h^{-1}$ kpc.  This is contradictory to the central cusp predicted by
the CDM numerical simulations \citep[][hereafter
NFW]{Navarro_etal_1996} and thus has motivated investigations of
different types of dark matter such as self-interacting dark matter
\citep{Spergel_Steinhardt_2000,
Moore_etal_2000}. \cite{Broadhurst_etal_2000} measured the arc
redshift to be $z_{\rm arc}=1.675$ and also built a lens model in a
simplified manner assuming NFW profiles for cluster galaxies. Their
model needed to include only the brightest eight cluster galaxies to
reproduce the observed lensed configuration. They reached the
conclusion that, in contrast to \cite{Tyson_etal_1998}'s model, the
average mass profile is consistent with an NFW profile.
\cite{Shapiro_Iliev_2000} pointed out that the fit by
\cite{Broadhurst_etal_2000} implies a cluster velocity dispersion that
is much larger than the value measured for this cluster by
\cite{Dressler_etal_1999}.

The galaxy velocity dispersion of $\sigma\simeq 1200$ km/s was found
by \cite{Dressler_Gunn_1992} and \cite{Dressler_etal_1999} based on
the data of about 100 galaxies.  \cite{Czoske_etal_2001,
Czoske_etal_2002} constructed a galaxy catalog of the wide-field
spectroscopic survey of the cluster, and newly identified a foreground
and a background group of galaxies well aligned along the line of
sight.  They clearly demonstrated the bimodal distribution of galaxies
in the redshift histogram for 300 objects in the neighborhood of
CL0024+17. If restricted to the galaxies in the main component, the
velocity dispersion is $\sim 600$ km/s. Thus if interpreted as a value
for a relaxed cluster, the dynamical mass would have only a quarter of
that previously derived \citep{Schneider_etal_1986}.

X-ray emitting intracluster gas is an excellent tracer of the dark
matter potential. \cite{Soucail_etal_2000} performed a combined
analysis of the {\it ROSAT} and the {\it ASCA} data and estimated the
cluster mass within the arc radius under the assumption of hydrostatic
equilibrium (the X-ray mass, hereafter). They found that there is
about a factor of $\sim3$ discrepancy between the X-ray mass and the
strong lensing mass \citep{Tyson_etal_1998, Broadhurst_etal_2000}.
They compared the extrapolated X-ray mass to the weak lensing mass
within $3~h_{50}^{-1}$Mpc and found it is again lower by a factor of
$\sim 3$.  Because the {\it ROSAT} HRI image suggested an elongated
gas distribution \citep[see also][]{BOHRINGER_ETAL.2000}, they
considered that the discrepancy may be partly caused by the irregular
mass distribution.

The `mass discrepancy problem' between X-ray and strong lensing mass
estimation has been reported in many other lensing clusters and the
X-ray mass is systematically lower than the strong lensing mass by a
factor of 2--5 \citep{Escude_Babul_1995, Wu_Fang_1997, Ota_etal_1998,
Hashimotodani_1999}. A variety of possible explanations have been
proposed; non-thermal pressure support, effect of complex mass
distribution, etc. \citep[e.g.][]{Hattori_etal_1999}. However, there
is as yet no definitive explanation, and it is possible that some
important physical process may not have been considered in regard to
the central region of clusters.

For the specific case of CL0024+17, however, there were still large
measurement uncertainties in both the X-ray temperature and the image
morphology, due mainly to heavy contamination from a 
bright seyfert galaxy that lies near the cluster center.  Thus for the
cluster mass estimation the temperature determination is crucial.  The
high-resolution {\it Chandra} data eliminate the contamination
from point sources and enable us to determine the X-ray spectrum and
the spatial distribution of the cluster gas simultaneously. In this
paper, we report accurate measurements of the temperature and
the morphology with {\it Chandra}, from which we consider the 
discrepancy between the X-ray and the strong lensing masses.

We use $H_0=50$ km/s/Mpc and $\Omega_0=1$, and thus $1\arcmin $
corresponds to $383~h_{50}^{-1}$ kpc at $z=0.395$.  The quoted errors
are the 90\% confidence range throughout the paper, except where
noted. We use the solar abundance ratio of iron atoms to hydrogen
atoms, ${\rm Fe/H} = 4.68\times10^{-5}$ \citep{Anders_Grevesse_1989}.

\section{Observation and source detection}

\subsection{{\it Chandra} observation of CL0024+17}
We observed CL0024+17 with the {\it Chandra} Advanced CCD Imaging
Spectrometer (ACIS-S) detector on September 20, 2000. The target was
offset from the ACIS-S nominal aimpoint with a Y offset of $-1'.33$ in
order to avoid the outskirts of cluster diffuse emission falling in a
gap between the CCD chips. The CCD temperature was $-120 {\rm
^{\circ}C}$. The data were reduced using CIAO version 2.2 with CALDB
version 2.15. We removed periods of high background levels exceeding
$3\sigma$ above the mean quiescent background rates. The net exposure
time was 37121 sec (93.2\% of the total exposure).  To improve the
{\it Chandra} astrometry we used the Aspect Calculator and corrected
the aspect offsets.  In Figure \ref{fig:image}, we show the ACIS-S3
image in the 0.5--7 keV band. The strongest X-ray peak of CL0024+17 is
at 00:26:36.0, +17:09:45.9 (J2000) and extended emission is detected
out to $\sim2\arcmin$ in radius.  We refer to the X-ray peak as the G1
peak hereafter.

We planned to set the roll angle at $127\arcdeg$ with a torelance of
$\pm 8\arcdeg$ in order to place the position of the ``perturbation'',
indicated by the weak lensing analysis \citep{Bonnet_etal_1994}, on
the ACIS-I2 chip. The perturbation is about $6\arcmin.7$
north-northeast of the gravitational shear coherent to the CL0024+17
center. However the tolerance was exceeded by $8\arcmin.4$ and the
actual roll angle was $135.14\arcdeg$, and thus the perturbation was
outside the ACIS field of view.

\subsection{Source detection in the ACIS-S3}
We searched for point-like sources in the ACIS-S3 field with the
WAVDETECT algorithm with a significance threshold parameter of
$10^{-6}$ and detected 38 sources. In the following analysis these
point sources were excluded with a radius of 7 times the size of the
point spread function (PSF) at the source position. The PSF size is
defined as the 40\% encircled energy radius at 1.5 keV. We
cross-correlated the source positions with the {\it ROSAT} HRI and
found 5 sources in the ACIS-S3 field of view \citep[S1, S2, S3, S4, S5
in][]{Soucail_etal_2000} were consistent with each other within
$1\arcsec\sim3\arcsec$, which is smaller than the pointing accuracy of
the {\it ROSAT} HRI (1$\sigma$ error is typically 6\arcsec;
\cite{BRIEL_ETAL.1997}).  For S1, there were two optical counterparts
within the {\it ROSAT} error box ($16\arcsec\times16\arcsec$), which
are a foreground starforming galaxy and one of the cluster member
galaxies.  The {\it Chandra} imaging clearly resolved S1 into two
point sources, whose J2000 coordinates are respectively determined to
be ($\alpha$, $\delta$)=(00:26:31.691, +17:10:21.71) and
(00:26:31.101, +17:10:16.48).  These are cataloged by
\cite{Czoske_etal_2001} as \#282 ($z=0.2132$) and \#267 ($z=0.4017$)
respectively, thus by comparing the positions we confirmed that the
{\it Chandra} astrometry is accurate down to $0\arcsec.3$.

\subsection{MG0023+171}
The radio source MG0023+171 with $z=0.946$ \citep{HEWITT_ETAL.1987}
was covered with the ACIS-S0 chip during the observation.  MG0023+171
has two optical counterparts separated by $5\arcsec$ which have been
interpreted as gravitationally lensed images. The large separation
angle implies a large mass-to-light ratio for the lensing matter
however the lensing object has not yet been identified.  We searched
for X-ray emission from a ``dark object'', but did not detect any
object using WAVDETECT with the threshold parameter of $10^{-5}$.  The
photon counts in the 0.5--7 keV band within a circle of radius
$5\arcsec$ centered at image A of MG0023+171 system is 6, without
subtracting background. We thus constrained the $3\sigma$ upper limit
on the X-ray energy flux as $8.3\times10^{-15} {\rm erg/s/cm^2}$
assuming a power-law spectrum with the photon index of 2.0. On the
other hand if we extrapolate the B magnitude of the radio source,
21.9, utilizing a typical $\alpha_{\rm OX}$ index for radio-loud
quasars of 1.6 \citep{Green_etal_1995}, the X-ray flux at 2 keV is
expected to be $6.4\times10^{-16} {\rm erg/s/cm^2/keV}$. We found that
this is below the detection limit of the current observation. If we
attribute the entire X-ray flux to a lensing object at an assumed
redshift of 0.4, the upper limit of the X-ray luminosity is
constrained to be $\sim 7\times10^{42}$ erg/s. Thus a massive object
such as a galaxy group or a galaxy cluster is unlikely to be the dark
lens candidate as long as one assumes the normal X-ray properties.

\section{Spectral analysis}\label{sec:spec}
\subsection{Overall spectrum}
We extracted the cluster spectrum from a circular region of $r
=1\arcmin.5$, centered on the G1 peak. The background was estimated
from the $2\arcmin.5 < r < 3\arcmin.2$ ring region and subtracted from
the above spectrum.  Note that we investigated the positional
dependency of the background by comparing two spectra accumulated in
the corresponding detector regions for the source and the background
spectra of the blank-sky observation data.  We found that the
difference of the background spectra from the two regions is
sufficiently small (the difference in intensity is less than 20\% in
each spectral bin and the overall normalization deviates by 4\%) and
the results of the present analysis are not affected within the
statistical errors. We generated the telescope response file (i.e. the
ARF file) for an extended source with the WARF procedure, which sums
the ARFs according to the weight of counts in each bin in the given
image region.  We found that the weighted ARF differs only by 2\% at
most in the ACIS-S energy band from that made for a point source.
Thus effect of telescope vignetting is negligible in comparison to the
statistical errors of the present data.

We fitted the cluster spectrum in the energy range of $0.5-7$ keV with
the MEKAL thin-thermal plasma model \citep{MEWE_ETAL.1985,
MEWE_ETAL.1986,KAASTRA.1992, LIEDAHL_ETAL.1995}. If we include the
hydrogen column density of the neutral absorption as a free parameter,
we obtained $N_{\rm H} = (5.3\pm2.5)\times10^{20}~{\rm
cm^{-2}}$. This is consistent with the Galactic value, thus we fixed
it at $N_{\rm H}=4.2\times10^{20}~{\rm cm^{-2}}$
\citep{DICKEY_ETAL.1990}. The result of the fit is shown in Figure
\ref{fig:spectrum} and Table \ref{tab:spec}.  This provided a good fit
to the data and the resulting $\chi^2$ value was 47.5 for 45 degrees of
freedom.  The temperature was determined to be $kT = 4.47$ keV with
the 90\% error range of $3.93-5.30$ keV. The result is consistent with
the previous measurement with {\it ASCA}, which had a
large ($\sim50$\%) uncertainty \citep{Soucail_etal_2000}.

We detected strong line emission at about 4.8 keV in our reference frame,
which is consistent with a redshifted, highly ionized Fe-K line
emitted from an object at $z=0.395$.  This is the first detection of
the iron line from CL0024+17. We obtained the metal abundance to be
$0.76^{+0.37}_{-0.31}$ solar (90\% error).  In the {\it Chandra}
spectrum there is also a significant contribution from the Fe-L complex
below 1 keV.  If we mask the energy bins between 4 keV and 5.2 keV
that cover the energy of Fe-K line, the MEKAL model fitting resulted
in the metal abundance of $1. 38^{+1.63}_{-0.65}$ solar (90\% error).
The best-fit value is higher than that we obtained from the above fit
however still consistent within the error ranges.  Thus the
metallicity for CL0024+17 is about a factor of 2 higher than those of
typical distant clusters \citep{Mushotzky_Loewenstein_1997,
MATSUMOTO_ETAL.2000, Ota_2001}.  On the other hand, the previous {\it
ASCA} spectral analysis by \cite{Soucail_etal_2000} did not constrain
iron emission lines because the photon statistics of the cluster
spectrum were seriously limited due to the contamination from the S1
emission.  The high metallicity value we derived is discussed later in
the paper (See \S\ref{subsec:xrayproperties}).  The X-ray luminosity
in the $0.5-7$ keV is $3.4\times10^{44}$ erg/s. The bolometric
luminosity is estimated to be $5.1\times10^{44}$ erg/s.

\subsection{Radial temperature distribution}\label{subsec:tprof}
In order to investigate the radial temperature profile, we accumulated
spectra from four annular regions centered on the G1 peak.  The radius
ranges were chosen so that each spectrum contains more than 400
photons. We fitted the spectra with the MEKAL model to determine the
temperatures, where the neutral absorption was fixed at the Galactic
value and the metallicity of the gas was allowed to vary.  In Figure
\ref{fig:tprof}(a), we plot the radial temperature profile and
$1\sigma$ error bars.  The temperature in each radial bin was
constrained with $13\sim20$\% accuracy.  We found that there is no
meaningful temperature variation with radius.

We further restrict the spectral regions to $8\arcsec$ and $4\arcsec$
(= 51 and 26 $h_{50}^{-1}{\rm kpc}$) from the G1 peak to constrain the
X-ray emission from G1.  From the MEKAL model fitting to the spectra,
we obtained the X-ray temperature to be 3.7 (2.7--5.5) keV and 3.4
(2.6--4.6) keV for the regions of $r<4\arcsec$ and $8\arcsec$,
respectively. Thus the temperature is still higher than 2.6 keV (90\%
confidence) at the center.

From the above analysis, we found that the intracluster gas is
consistent with being isothermal out to $\sim600~h_{50}^{-1}$
kpc. Note that the fraction of the maximum radius to the virial
radius, $r_{200}$ (See \S\ref{subsec:xrayproperties} for definition)
is 0.41. In Figure \ref{fig:tprof}(b), we show the 68\% confidence
contours for the temperature and the metallicity for four annular
regions obtained from the MEKAL model fitting. We found that the
best-fit metallicity at the cluster center is as high as 1 solar,
however, due to the large statistical errors the abundance gradient is
not significant.

\section{Image analysis}\label{sec:image_analysis}
\subsection{X-ray surface brightness and galaxy distribution}
Though the original ACIS CCD has a pixel size of $0\arcsec.492$, we
rebinned the image increasing the bin dimensions by a factor of 4.
Thus 1 image pixel is $1\arcsec.968$ which is 12.6 kpc at the
cluster's frame.  We restrict the energy range to 0.5 -- 5 keV in the
image analysis in order to avoid the hard energy band where the
background dominates the total spectrum. We find that there is a
second X-ray peak at 00:26:35.1, +17:09:38.0 (J2000). We refer to the
second X-ray peak as G2 hereafter. The G2 peak is about 100 kpc
southwest of the G1 peak. We investigated the correlation between the
X-ray surface brightness and the member galaxies cataloged by
\cite{Czoske_etal_2001}.  As shown in Figure \ref{fig:2dfit}a, by
superposing the galaxies with redshift range of $0.38 - 0.41$ and V
magnitude smaller than 22 on the {\it Chandra} X-ray image, we
recognized that the three central bright elliptical galaxies are
located at positions that coincide with the G1 and the G2 X-ray
peaks. The object numbers and the redshifts cataloged by
\cite{Czoske_etal_2002} are \#380 ($z=0.3936$), \#374 ($z=0.3871$),
and \#362 ($z=0.3968$) from east to west in Figure \ref{fig:2dfit}a.
Note that G1 contains \#380 and \#374 however the X-ray peak position
is more consistent with \#380 (Figure \ref{fig:2dfit}a).

\subsection{1-D fitting of the X-ray surface brightness distribution}\label{subsec:1dfit} 

We investigate the X-ray surface brightness distribution by
1-dimensional and 2-dimensional model fitting in this and the next
subsection, respectively. The latter is more precise in the sense that
the central position of the X-ray emission can be included as a model
parameter. Here we derive an azimuthally averaged surface brightness
distribution centered at the strongest peak, G1, from the 0.5--5 keV
image and evaluate the overall shape by fitting a $\beta$-model,
$S(r)=S_0 ( 1+(r/r_c)^2 )^{-3\beta+1/2}$
\citep{Cavaliere_Fusco_1976}. The telescope vignetting was corrected
by dividing the image with the exposure map. The G2 peak was excluded
with a circular region of $4\arcsec$ in radius, which covers more than
95\% of the G2 emission, when calculating the profile.

We found that a single-component $\beta$-model is rejected at the
99.1\% level and there is clearly strong excess emission over the
model in the innermost region (Figure \ref{fig:1dfit}a).  We also
notice that there is a point of inflection in the profile at around
$r\sim 100$ kpc, where systematic variation of the residuals are
clearly visible.  This leads us to test the hypothesis that the
surface brightness consists of two components. If we introduce a
double-$\beta$ model consisting of two $\beta$-profiles with different
core radii, the fit was significantly improved from the single $\beta$
model and gave $\chi^2/dof= 183.2/194$.  In Figure \ref{fig:1dfit}b,
we show the results for the double $\beta$-model fitting. The
derived parameters are listed in Table \ref{tab:1dfit}.  Note that the
background constant was included as a free parameter in the fit and
determined to be $C=8.5^{+0.3}_{-0.2}\times10^{-9}~{\rm
  counts/sec/cm^2}$. A strong degeneracy exists between parameters
$r_{c, in}$ and $\beta_{in}$ (i.e. the core radius and the slope for
the inner component of the double $\beta$-model) that prevents the fit
from converging properly. We tried different fixed values of $\beta_{in}$
ranging between 0.5 and 1.5 and found that all fitting parameters
except for $r_{c, in}$ are consistent with the result for
$\beta_{in}=1$ within the 90\% error bars. As long as $0.6 \le
\beta_{in} \le 1.5$ the resultant $r_{c, in}$ value is also
statistically consistent with the result for $\beta_{in}=1$. Thus we
chose to fix $\beta_{in}$ to 1.

We also tested a model for gas distribution in the case that the gas
is in hydrostatic equilibrium in the NFW potential
\citep{Navarro_etal_1996}. \cite{Suto_etal_1998} provided a useful
fitting formula for the X-ray surface brightness distribution in the
generalized form of the NFW-type potential, $\rho(x)\propto
1/(x^{\alpha}(1+x)^{3-\alpha})$, which gives a good fit for
$1.0\le\alpha \le 1.6$. We refer to the formula \citep[i.e. equation
29--32 in][] {Suto_etal_1998} as the NFW-SSM model hereafter. It is
also known that the gas density profile for the NFW model is well
approximated by the $\beta$-model \citep{MAKINO_ETAL.1998}, though it
has a slightly steeper slope in the innermost region. Thus it is worth
quantifying the parameter $\alpha$. We first fitted the radial surface
brightness distribution with the NFW-SSM model for $\alpha=1$
(i.e. the NFW case). The result is presented in Table
\ref{tab:1dfit}. We found that it is not accepted at the 90\%
confidence level. If the $\alpha$ value is allowed to vary within the 
range 1.0 to 1.6, we found that the $\alpha=1.6$ case
resulted in the minimum $\chi^2$ value of 199.8 for 196 degrees of
freedom. Thus the central X-ray profile appears to be much
steeper than that expected from the original NFW profile. However the
model with $\alpha=1.6$ is unlikely because $r_s$ became
unrealistically large (more than one order of magnitude larger than
the value for the $\alpha=1$ case).

Thus from the 1-dimensional analysis, the X-ray spatial distribution
is not sufficiently described either by the single-component
$\beta$-model or by the NFW-SSM model and is significantly better
fitted with the double-$\beta$ model.

\subsection{2-D fitting of the X-ray surface brightness distribution}

In order to determine the X-ray emission profile of the ICM more
precisely, we fitted the 2-dimensional surface brightness distribution
with a model consisting of three $\beta$ profiles which we consider to
represent emission from two compact components associated with the G1
and the G2 peaks and ``cluster'' component and the constant
background;
\begin{equation}
S(r)=\sum_{i=1}^{3} S_{0,i} ( 1+(r/r_{c,i})^2 )^{-3\beta_i+1/2} + C. 
\label{eq:threebeta}
\end{equation}
We fitted the image of the central $(100~{\rm image\;pixel})^2 =
(3\arcmin.3)^2$ region using the maximum-likelihood method.  The
center positions of the first two compact components were fixed at the
G1 and the G2 peaks, respectively, while for the third component,
which we consider describes the ICM emission, the position was allowed
to vary.  Because the slope parameters for the G1 and the G2
components were insensitive to the fit, we first assumed King profiles
\citep{KING.1962}, namely $\beta$ models with $\beta_1 = \beta_2 = 1$.
The core radius of G2 was not resolved with the spatial resolution of
the current image analysis, thus fixed at $r_2 = 10$ kpc as is typical
for an elliptical galaxy.  The background level, $C$, was fixed at the
value that was estimated from the 1-D analysis, $8.5\times10^{-9}~{\rm
counts/sec/cm^2}$.  We performed the fit with the SHERPA package,
where we included the exposure map to convolve the model image with
the telescope and detector responses. The exposure map was calculated
at an energy of 0.8 keV, representative of the spectral energy
distribution. The results of the fits are shown in Table
\ref{tab:2dfit} and Figure \ref{fig:2dfit}. In order to check the
goodness of the fit, we rebinned the image into two single dimensional
profiles in the right ascension and declination directions (Figure
\ref{fig:2dfit}d) and calculated the $\chi^2$ values between the model
and data profiles to find they are sufficiently small ($\chi^2/dof
=112.4/99$ and 108.2/99 for the $\alpha$- and $\delta$-directions,
respectively). The best-fit cluster center position is significantly
shifted from the G1 peak by $12\arcsec.5$, that is $80~h_{50}^{-1}$kpc, 
towards the southwest direction. There is no optical counterpart
present at this position.

In Table \ref{tab:2dfit}, the results of 2-dimensional image fitting
are listed.  For G1 we obtained a core radius of $52^{+11}_{-9}$
kpc. The cluster component was found to be much more extended than the
G1 component, being described by a spherical $\beta$-model with
$r_{c,3}= 210^{+33}_{-30}$ kpc and $\beta_3 = 0.71^{+0.07}_{-0.06}$.
The 90\% confidence contour of the $\beta_3$ and $r_{c,3}$ are shown
in Figure \ref{fig:contour}. These two values are consistent with
those we obtained from the 1-D fitting even though the cluster center
is significantly displaced from that assumed in the 1-D analysis. We
consider that this is because the displacement is small in comparison
to the cluster core radius, inside which the surface brightness is
flat. The small core radius of the G1 emission is consistent with the
result from the {\it ROSAT} HRI though it gave a smaller $\beta$
value. In fact we obtained a similarly small $\beta$ in the 1-D
analysis (see \S\ref{subsec:1dfit}). This may be naturally explained
by the existence of emission from the larger core component. In
addition, though we fixed the $\beta_1$ value at 1 for G1 because the
inner slope parameter is not sensitive to the fit, as for the 1-D
fitting, we confirmed that the resultant $\beta$-model parameters for
all three components do not change within statistical errors if $0.6
\le \beta_{1} \le 1.5$.

We estimated the luminosities for the three components by dividing the
total luminosity obtained from the spectral fit according to the ratio
of photon counts between the three.  Here we assumed isothermality for
the three components. The results are also shown in Table
\ref{tab:2dfit}.  The luminosity of the ``cluster'' component is
$4.5\times10^{44}$ erg/s and dominates the total luminosity.  We
consider that the emission of G2 can be attributed to an elliptical
galaxy because our derived luminosity is within the scatter of other
ellipticals \citep{FABBIANO_ETAL.1992, MATSUSHITA.2001} and there is a
good positional coincidence with the member elliptical galaxy, \#362.
On the other hand the luminosity of the G1 component is higher for a
typical elliptical galaxy and the core radius is also comparable to
that of other cluster with a small core radius \citep{OTA_ETAL.2002}.
We will discuss the properties of the G1 component in more detail
later.

Furthermore, we tested the significance of the ellipticity of the
cluster image by substituting the third component in equation
\ref{eq:threebeta} for the elliptical $\beta$-model. We found the
ellipticity is not significant; the 90\% upper limit was $\epsilon <
0.2$. On the other hand, \cite{Soucail_etal_2000} and
\cite{BOHRINGER_ETAL.2000} reported that the HRI image is slightly
elongated in a northeast-southwest direction and can be fitted with a
ellipse with an ellipticity of $\sim
0.2-0.3$. \cite{Soucail_etal_2000} also mentioned that there is a
significant twist of the position angle at about 100 kpc from the
center.  We consider that the significant shift of the center position
of cluster emission by 80 kpc and the presence of the second peak at
100 kpc off of the strongest peak in the {\it Chandra} image can
account for the moderate ellipticity and the twist derived by the
elliptical model analysis with the HRI.

\section{Mass estimation and comparison} 

In this section we will derive the projected cluster mass profiles
based on the results of the X-ray analysis and estimate the cluster
mass enclosed within the arc radius.  We will directly compare these
with the results from the lensing mass reconstruction.

\subsection{X-ray mass for the $\beta$-model}\label{subsec:xraymass}
From the spectral and spatial analysis with {\it Chandra}, we found
that the cluster gas is consistent with being isothermal and can be
described with the spherical $\beta$-model after removing the local
emission of G1 and G2. In what follows, we regard the ``cluster''
component, $i=3$ in the three $\beta$-model fit, as diffuse emission
coming from the ICM confined in the cluster dark matter
potential. Since the G1 core radius is much smaller than the arc
radius ($r_{c,1} \ll r_{\rm arc}$), while the cluster core radius
$r_{c,3}\sim r_{\rm arc}$, the G1 component is thought to be a minor
contributor to the X-ray mass estimation. The effect of including 
the G1 mass will be discussed later in this section.

Assuming the gas is isothermal and spherically distributed and in
hydrostatic equilibrium, the density of matter at a radius $r$ is
estimated from the $\beta$-model as
\begin{equation}
\rho_{\beta}(r) = \frac{3kT\beta}{4\pi r^2G\bar{m}}
\left[\frac{3r^2}{r^2+r_c^2}-\frac{2r^4}{(r^2+r_c^2)^2}\right].
\end{equation}
The projected X-ray mass density profile is given by
\begin{equation}
\Sigma_{{\rm X},\beta}(R) = \int_{-\infty}^{\infty} 
\rho_{\beta}(\sqrt{R^2 + z^2}) dz =
{3kT\beta \over 4G\bar{m}}
\left[{R^2+2r_c^2\over (R^2+r_c^2)^{3/2}}\right].\label{eq:massdensity}
\end{equation}
In Figure \ref{fig:densityprof} we show the mass density profile
calculated for $\beta$-model.  The cylindrical cluster mass within a
certain radius $R$ is calculated by integrating the projected density
profile, which can be written as follows \citep{Ota_etal_1998}.
\begin{equation} M_{{\rm X},\beta}(R) = 
\frac{3kT\beta}{G\bar{m}}\frac{\pi}{2}\frac{R^2}{\sqrt{R^2 + r_c^2}}
\label{eq:xraymass}
\end{equation} 
Then from the results of 2D fitting for the cluster 
component we estimated the X-ray mass within the arc radius of $R_{\rm 
arc}=35\arcsec=220~h_{50}^{-1}$ kpc to be
\begin{equation} 
M_{{\rm X},\beta}(220~h_{50}^{-1} {\rm kpc}) 
= 0.84^{+0.20}_{-0.13}\times10^{14}~h_{50}^{-1}{\rm M_{\sun}}.
\label{eq:mx_beta}
\end{equation} 
Because the statistical errors of $\beta$ and $r_{c}$ are coupled
with one another (Figure \ref{fig:contour}) though $kT$ is
independently determined from the $\beta$-model parameters, we
determined the error associated with the X-ray mass by evaluating it
in the statistically allowed domain of the three dimensional parameter
space.

On the other hand, adopting the arc redshift of $z_{\rm arc}=1.675$,
the strong lensing mass is estimated to be
\begin{equation} 
M_{\rm lens} (<214~h_{50}^{-1} {\rm kpc})=(3.117\pm0.004)\times 
10^{14} ~h_{50}^{-1}{\rm M_{\sun}}
\label{eq:lensmass_tyson}
\end{equation} 
from equation (2) in \cite{Tyson_etal_1998}.
\cite{Broadhurst_etal_2000} derived the strong lensing mass as $M_{\rm
lens}(<200~h_{50}^{-1} {\rm kpc})=(2.22\pm0.06)\times10^{14}~
h_{50}^{-1}{\rm M_{\sun}}$, which is smaller by about 30\% than
equation \ref{eq:lensmass_tyson}. However, comparing $M_{\rm X,\beta}$
to $M_{\rm lens}$, a mass discrepancy of a factor of 3 is evident.

As shown in Figure \ref{fig:densityprof}, since the surface mass
density of the cluster component obtained by the isothermal
$\beta$-model is less than the critical surface mass density,
$\Sigma_{\rm crit}=(c^2/4\pi G)/(D_s/D_d D_{ds})=2.1\times10^3{\rm
M_{\sun}/pc^2}$ everywhere, the existence of the gravitationally
multiple images in this cluster means that the mass distribution of
the cluster is far from the isothermal $\beta$-model and/or the
existence of an extra mass component along the line of sight toward
the cluster central region is required.  

\subsection{X-ray mass for the NFW-model}
The X-ray surface brightness profile of the NFW potential
\citep{Navarro_etal_1996} is similar to that of the $\beta$-model and
can be converted from the $\beta$-model parameters through the
relations of $r_s=r_c/0.22$ and $B=15\beta$
\citep{MAKINO_ETAL.1998}. We thus derived the NFW density profile
based on the results of $\beta$-model fitting,
\begin{equation}
\rho_{\rm NFW}(r) = \frac{\rho_s r_s}{r (1+\frac{r}{r_s})^2}, 
\end{equation} 
where $\rho_s = kT B/4\pi G \bar{m} r_s^2$. We then calculated the
projected mass density profile in the same manner as equation
\ref{eq:massdensity}.  The mass density profile is much steeper at the
innermost region than that of the $\beta$-model and the central
surface mass density can be higher than the critical surface mass
density, however, those two models are not distinguishable at the arc
radius (Figure \ref{fig:densityprof}). We estimated the X-ray mass to
be
\begin{equation}
M_{\rm X, NFW}(220~h_{50}^{-1} {\rm kpc})=
0.74^{+0.18}_{-0.11}\times10^{14}~h_{50}^{-1}{\rm M_{\sun}}. 
\end{equation}

This shows that the NFW model does not improve the enclosed mass
within the arc radius. Therefore, adopting the NFW model can not be
the main solution for the mass discrepancy problem of this cluster.

\subsection{Estimation of mass associated with G1}\label{subsec:G1mass}

We will estimate the projected mass associated with the G1 component
under some simple assumptions for the purpose of constraining
its possible contribution to the X-ray mass estimate.  Firstly, we
consider the mass of the two bright elliptical galaxies inside
G1. Their velocity dispersions and effective radii were measured to be
$\sigma=317$ km/s and $r_e=5.9 ~h_{50}^{-1}{\rm kpc}$ for \#380 and
$\sigma=275$ km/s, and $r_e = 3.4 ~h_{50}^{-1}{\rm kpc}$ for \#374,
respectively \citep{VANDOKKUM_ETAL.1996}. Then calculating the
projected lensing mass under the Singular Isothermal Sphere model, we
obtain $M_{\rm SIS} \sim \pi \sigma^2/ (2 G r_e)=4.4\times10^{11}~{\rm
M_{\sun}}$ for \#380 and $M_{\rm SIS}=1.9\times10^{11}~{\rm M_{\sun}}$
for \#374. Thus the sum of the two galaxies can increase $M_{\rm X}$
by only 1\%.

Second, we will treat the G1 component as the cluster scale
substructure embedded in or projected on the cluster.  One reason in
support of the idea is that the observed X-ray properties such as the
temperature, the luminosity and the core radius all exceed those
observed for a typical elliptical galaxy and they are rather close to
a single cluster. However there is no unique way to evaluate this
component in the $M_{\rm X}$ estimation, as it is not clear how this
is physically related to the surrounding ICM. Thus our estimate of the
G1 mass supposes that it is locally in hydrostatic equilibrium and
dominates the mass out to a certain cutoff radius ($r_{\rm cut}\sim
100 h_{50}^{-1} {\rm kpc}$) at which the inflection in the X-ray
surface brightness is found.  Thereby we obtain $M_{\rm G1}(100~
h_{50}^{-1}{\rm kpc}) =
0.51^{+0.20}_{-0.14}\times10^{14}~h_{50}^{-1}{\rm M_{\sun}}$ utilizing
equation \ref{eq:xraymass} and the measured temperature within a
radius of $8\arcsec$ circle centered on G1.  Adding $M_{\rm G1}$ to
equation \ref{eq:mx_beta}, we obtain $M_{{\rm X},\beta}\lesssim
1.7\times10^{14}~h_{50}^{-1}{\rm M_{\sun}}$. We consider this will
give a secure upper limit of the X-ray mass estimation under the
current isothermal $\beta$-model analysis.  The total gas mass
associated with the G1 clump within a sphere of $r=r_{\rm cut}$ is
$M_{\rm gas, G1}(100~ h_{50}^{-1}{\rm kpc}) = 5.7^{+1.8}_{-1.5}\times
10^{11}h_{50}^{-5/2}{\rm M_{\sun}}$ and has a negligible contribution
to the total gravitating mass.  Thus this upper limit is still
significantly smaller than $M_{\rm lens}$ by a factor of $\sim 2$. If
we compare to the Broadhurst et al.'s mass, the discrepancy reduces to
about 30\%.

We drew critical lines of the lens model for a source at $z=1.675$
where the cluster component and the G1 component are included (Figure
\ref{fig:critical_line}).  The surface mass distribution of the G1
component is described by the equations (2) and (3) with above
parameters for $r<r_{\rm cut}$ where $r$ is the projected radius from
the G1 center.  The surface mass density beyond $r_{\rm cut}$ is
assumed to be zero. Other parameters are the same as in
\S\ref{subsec:xraymass}.  Inclusion of the G1 component turns the
isothermal $\beta$-model into the supercritical and a tangential
critical line appears.  Thus G1 has a significant contribution to the
lensing effect. Moreover we notice that $r_{c,1}$ is fairly close to
the core size of the dark matter distribution found in
\cite{Tyson_etal_1998}. We then consider that this also assures the
significance of the existence of the G1 potential. If the lens model
is able to explain the observed multiple images, the image appearing
on the same side of the central image, which is in the south-east side
from the cluster center (i.e. image B in Figure
\ref{fig:critical_line}), should be enclosed by the tangential
critical line.  However, the inclusion of the G1 component described
above is not enough to explain the observed lensed multiple images
because the tangential critical line can not reach the south-east
image.

\subsection{Cluster center position}

Next we directly compare the cluster center position derived by the
X-ray image analysis to that of dark matter profile modeled by
\cite{Tyson_etal_1998}. We found that their center coordinates,
(00:23:56.6, +16:53:15) (1950) are shifted if comparing their Figure 2
to the latest optical coordinates of central bright ellipticals from
\cite{Czoske_etal_2001}. Thus utilizing the coordinates of galaxy
\#380 in \cite{Czoske_etal_2001} and the relative distance from \#380
galaxy to the DM center, $1\arcsec.5$ to the east and $4\arcsec.5$ to
the south, the DM center coordinates correspond to about (00:26:35.8,
+17:09:39.4) in J2000.  Thus we found that the X-ray center position
(00:26:35.6, +17:09:35.2) is shifted to the south-southwest by
$4\arcsec$ from the dark matter center. If we take into account the
measurement uncertainties of $\sim 2\arcsec$, we consider that the
shift is marginal. Moreover \cite{Tyson_etal_1998} derived the maximum
point of diffuse intracluster light that is not associated with
visible galaxies and showed it is displaced $3\arcsec$ west-southwest
from the DM center. The X-ray center position is closer to this than
the DM center.

\section{Discussion}

\subsection{Properties of the intracluster gas}\label{subsec:xrayproperties}

We will discuss the X-ray properties of the gas focusing on the X-ray
luminosity-temperature ($L_X-T$) relation, gas-mass fraction and the
metallicity based on the {\it Chandra} analysis.

The location of the cluster on the $L_X-T$ plane will provide
important information for understanding the physical status of the ICM
in the cluster central region. The observational data of clusters are
known to exhibit a large scatter around the best-fit power-law model.
\cite{FABIAN_ETAL.1994} noted that the $L_X-T$ relation is roughly
divided into two sequences whether the clusters are XD or non-XD.  The
XD cluster is the cluster in which the X-ray emission is highly
concentrated in the central giant elliptical galaxy and the X-ray peak
position coincides with the center of the central galaxy.  In the
non-XD cluster, the emission is diffuse and the X-ray peak position is
largely offset from the central giant galaxy.  \cite{ARNAUD_ETAL.1999}
derived the $L_X-T$ relation with the analysis sample restricted to
non-XD clusters and showed a smaller scatter. A different approach was
taken by \cite{OTA_ETAL.2002}, concerning the cluster core radius.
They performed a systematic analysis of 79 distant clusters with the
{\it ROSAT} HRI and {\it ASCA} to study the X-ray structure of
clusters in $0.1<z<1$. They determined the average X-ray temperatures
and the bolometric luminosities with {\it ASCA} and the X-ray surface
brightness distributions with the {\it ROSAT} HRI by utilizing the
isothermal $\beta$-model, and found there is not any significant
redshift dependence in the X-ray parameters including the temperature,
$\beta$-model parameters, and the central electron density. They
discovered that the distribution of the core radius shows distinct two
peaks at 60 kpc and 220 kpc. If dividing the cluster samples into two
subgroups corresponding to the two peaks in the core radius
distribution, they show differences in the X-ray and optical
morphologies and in the $L_X-T$ relation. In particular, the
normalization factor of the $L_X-T$ relation significantly differs
according to the core sizes: at a certain temperature, the luminosity
is higher for a cluster with small core radius and approximately $L_X$
follows $r_c^{-1}$. From these observational results, they suggested
that the clusters are divided into at least two subgroups according to
the core radius. In \S\ref{sec:image_analysis}, we showed that the
X-ray spatial distribution of CL0024+17 is described with a
superposition of two $\beta$-model components with $r_{c,1}\sim 50$
kpc and $r_{c,3}\sim210$ kpc.  Therefore, the main cluster component
is classified as a non-XD cluster or a large core radius cluster,
while the G1 component is classified as an XD or a small core radius
cluster.

We will compare our result for CL0024+17 with \cite{ARNAUD_ETAL.1999}
and \cite{OTA_ETAL.2002}. First, employing the $L_X-T$ relation of
\cite{ARNAUD_ETAL.1999}, the luminosity is expected to be
$\log{L_X}=44.69\pm0.05$ for the observed temperature of 4.5 keV. Thus
the X-ray luminosity of CL0024+17 determined with {\it Chandra},
$\log{L_{{\rm X, bol}}}= 44.7$, is in a good agreement.  
Moreover, we compare our result to the $L_X-T$ relation for clusters
with large($r_c>135$ kpc) core, derived by \cite{OTA_ETAL.2002} since
the main cluster component whose core is $r_{c,3}\sim210$ kpc is
responsible for the cluster luminosity (See Table \ref{tab:2dfit}). We
found that the luminosity of CL0024+17 is smaller than the other
distant cluster samples with similar temperatures, however, it is
within a scatter of the data points and is quite consistent with the
mean relation of the large core ($r_c>135$ kpc) clusters. Therefore in
terms of the $L_X-T$ relation, the X-ray emission of CL0024+17 agrees
with other clusters that do not have strong central X-ray emission
dominating the total luminosity.  Adopting the estimated G1 luminosity
of $5.5\times 10^{43}{\rm erg/s}$ (Table \ref{tab:2dfit}), the
temperature of 1.2 keV is obtained from the $L_X-T$ relation for the
small core radius clusters. The obtained temperature is significantly
lower than that constrained by the X-ray observation
(\S\ref{subsec:tprof}).  This might be the signature that the G1
component is now in the process of merging with a cluster or has
undergone a recent merger and therefore the characteristics of the G1
component can not be estimated by using a relation for relaxed
clusters.

Since the observed X-ray spectrum is well fitted with the
single-temperature MEKAL model (\S\ref{sec:spec}), the contribution of
additional cool emission is suggested to be not important in this
cluster. We can then estimate the time scale of radiative cooling at
the cluster center. By deprojecting the central surface brightness of
the $\beta$ profile, we obtained the central electron density for the
cluster component as $n_{e0,3}=(5.8\pm0.6)\times10^{-3}~{\rm
cm^{-3}}$. Then the cooling time scale is 11.4 Gyr at the center.
This is longer than the age of the Universe, 7.9 Gyr. Thus radiative
cooling is not likely to be effective. On the other hand, the electron
density is higher for G1, where
$n_{e0,1}=(2.5\pm0.5)\times10^{-2}~{\rm cm^{-3}}$. This yields a much
shorter cooling time of 2.8 Gyr. Thus cooling may occur very
effectively in the G1 region. However it is contradictory that we did
not find a significant temperature decrement in the spectral analysis
(See \S\ref{subsec:tprof}).  A similar situation has been found in
some nearby clusters with short cooling timescales
\citep[e.g.][]{Tamura_etal_2001} and other possibilities to prevent
cooling have been proposed \citep[e.g.][]{Bohringer_etal_2002}.  We
will discuss the possibility of a two-phase state of the ICM in the
next subsection in detail.


We estimate the gas-mass fraction in the cluster within the virial
radius in order to compare it with the mean baryon fraction in the
Universe determined by the Wilkinson Microwave Anisotropy Probe (WMAP)
experiments \citep{Spergel_etal_2003}. In this analysis, we use the
standard set of the cosmological parameters determined by WMAP,
$\Omega_m=0.3$, $\Omega_{\Lambda}=0.7$, and $h=0.7$.  The virial
radius is defined as the radius at which the cluster density is equal
to 200 times the critical density of the universe at the cluster
redshift, $r_{200}=1.39~h_{0.7}^{-1}$ Mpc. Then the virial mass (the
spherical hydrostatic mass), gas mass and gas-mass fraction calculated
at $r_{200}$ under the $\beta$-model are
\begin{eqnarray}
M_{200} &=& 4.6^{+1.4}_{-0.9}\times10^{14}~h_{0.7}^{-1}{\rm M_{\sun}},
\label{eq:m200}\\ M_{\rm gas} &=&
6.5^{+0.6}_{-0.6}\times10^{13}~h_{0.7}^{-5/2}{\rm M_{\sun}},\\ f_{\rm gas}
&=& M_{\rm gas}/M_{200} = 0.14^{+0.05}_{-0.04}~h_{0.7}^{-3/2}\label{eq:fgas}.
\end{eqnarray}
The gas-mass fraction at the virial radius is consistent with the
Universal baryon fraction derived by WMAP, $\Omega_b/\Omega_m=0.16$,
within 35\% accuracy.  Thus we consider the virial mass to be properly
evaluated with $M_{200}$ and the X-ray temperature to represent the
virial temperature of the cluster potential.

As shown in the spectral analysis, the average iron abundance is about
twice as high as the typical value. Since the metallicity is a
quantity defined relatively to the amount of hydrogen, there are two
situations that will explain this high value: a medium rich in iron or
a medium poor in hydrogen. Because the gas-mass fraction (Equation
\ref{eq:fgas}) is comparable to that observed in other clusters
\citep{MOHR_ETAL.1999, Ota_Mitsuda_2001}, it is plausible that the ICM
was highly enriched in iron. As for nearby clusters, the iron mass in
ICM, $M_{\rm Fe}$, is known to have a clear correlation with the
optical luminosity of E+S0 members \citep{Arnaud_etal_1992} and with
the total blue luminosity
\citep{Renzini_1997}. \cite{Schneider_etal_1986} mentioned that the
luminosity of CL0024+17 places it among the richest of any known
clusters. Because the previous measurement of the cluster metallicity
showed no significant redshift evolution
\citep[e.g.][]{Mushotzky_Loewenstein_1997}, and as the contribution of
the metal production by type Ia SNe after $z=0.4$ can be negligible
according to the model calculation by \cite{Mihara_Takahara_1994}, we
will compare the iron mass observed at $z=0.395$ to the optical
luminosity corrected for passive evolution in ellipticals below.
Utilizing the local relation, $M_{\rm Fe}\sim
1.6^{+3.5}_{-1.1}\times10^{-2} L_{\rm V,E+SO}^{1.0\pm0.30}$
\citep{Arnaud_etal_1992} and the visual luminosity of ellipticals
evolved to the present day $L_{V}^{\rm
E}(z=0)=1.2\times10^{12}h_{50}^{-2}{\rm L_{\sun}}$ at
$r=400\,h_{50}^{-1}$ kpc \citep{Smail_etal_1997}, the iron mass is
calculated as $1.9^{+4.2}_{-1.3}\times10^{10}h_{50}^{-2.5}{\rm
M_{\sun}}$, where we adopted the best-fit slope of the $M_{\rm}-L_{\rm
V,E+SO}$ relation, i.e. 1, and included only the $1\sigma$ error of
the normalization factor.  Though the relation of
\cite{Arnaud_etal_1992} was derived at a radius of
$3\,h_{50}^{-1}$Mpc, we assumed here that iron distribution is
proportional to that of the galaxy and then it is applicable for a
smaller radius.  On the other hand, the {\it Chandra} analysis yields
$M_{\rm Fe}(<400\,h_{50}^{-1}{\rm
kpc})=(2.5\pm0.2)\times10^{10}h_{50}^{-2.5}{\rm M_{\sun}}$ where we
adopted the mean metallicity measured within $600\,h_{50}^{-1}$ kpc,
0.76 solar. Because this is by about 30\% higher than the above
calculation, there may be an abundance variation from cluster to
cluster, as suggested from the recent {\it Chandra} results of
metallicity maps: the inhomogeneous metal distributions were found in
the core regions of some nearby clusters such as Perseus
\citep{Schmidt_etal_2002} and A2199
\citep{Johnstone_etal_2002}. However, since the iron mass of CL0024+17
is consistent with the $M_{\rm Fe}-L_{\rm V,E+SO}$ relation within the
1$\sigma$ errors, the variation is not statistically significant.  To
confirm this, we need future observations of the metallicity map with
higher sensitivities.

The observed high metallicity can be explained by metal ejection from
the elliptical galaxies.  Since the spiral galaxy fraction, $f_{\rm
sp}=0.40$, is not different from other clusters with a similar mass
\citep{Smail_etal_1997}, the above result indicates that very
effective galaxy formation had occurred in the central region of
CL0024+17. Besides, the optical observations showed that the spatial
distribution of galaxies is centrally concentrated and has a very
compact core \citep{Smail_etal_1996, Schneider_etal_1986}. Then
because of the high concentration of bright elliptical galaxies the
metal will also follow the concentrated distribution.  As mentioned
above, there may also be a complex metal distribution in the cluster
core. However due to the poor photon statistics, we are not able to
derive the 2-dimensional abundance map. For this reason we studied the
radially-averaged abundance profile in \S\ref{subsec:tprof}.  Our
result shown in Figure \ref{fig:tprof}b indicates that the metal
abundance is as high as $\sim 1$ solar at the innermost region and
then not in conflict with an abundance profile in which $Z_{\rm Fe}
\sim 1$ within the central $r\lesssim10\arcsec$ region and $Z_{\rm Fe}
\sim 0.3$ at the outer region.  In order to further constrain the
spatial distribution and the history of the metal production, better
photon statistics for the X-ray spectrum are required.

\subsection{Mass discrepancy problem}\label{subsec:massdiscrepancy}

We found from the {\it Chandra} observation that the X-ray surface
brightness distribution is mainly described by a superposition of two
extended components well fitted with spherical
$\beta$-profiles. Furthermore we do not find any significant
temperature structure. Thus it is likely that the gas is relaxed in
the cluster potential and, therefore, that hydrostatic equilibrium
will be a good approximation in the X-ray mass estimation. We further
discuss any possible cause of the mass discrepancy below.

As shown in the previous section, we can consider that the X-ray
temperature represents the virial temperature of the cluster component
since the gas mass fraction of the cluster component within the virial
radius is consistent with the universal baryon fraction obtained by
WMAP.  Therefore, the mass discrepancy reported in the previous
section is telling us that there is a lack of our current
understanding for the nature of the cluster central region.

In some nearby clusters, particularly XD clusters
\citep{Forman_Jones_1990}, an additional cool component is required to
explain the X-ray emission of the central $\sim100~h_{50}^{-1}$kpc
regions \citep{Makishima_etal_2001}. On the contrary
\cite{Makishima_etal_2001} also suggested that non-XD clusters appear
to be isothermal with little metallicity gradient toward the
center. Taking into account the fact that the current target cluster
is a non-XD cluster and that the total X-ray emission is not dominated
by the central G1 emission, our results shown in \S\ref{sec:spec} seem
to be more consistent with their picture of non-XD systems.  However
if the ICM of CL0024+17 is in a two-phase state and the thermal
pressure of the gas in each phase balances against gravitation, it
would greatly increase the X-ray mass estimation. Thus we attempted to
fit the spectra for both (1) the cluster region outside 100 kpc from
G1, i.e. $0\arcmin.26<r<1\arcmin.5$ and (2) the G1 region,
$r<0\arcmin.26$ with a two-component MEKAL model. Because we are not
able to constrain the two temperatures simultaneously under the
current photon statistics, we fixed the temperature and the
metallicity for the cool phase respectively at 1 keV and 1 solar while
the spectral normalization was adjustable. For the region (1), we
found that the flux of the cool phase is only about 4\% of the hot
phase and the temperature of the hot-phase is 5.5 (4.4 -- 7.2) keV,
which is consistent with the result of the single-phase model shown in
Table \ref{tab:spec} within errors. Thus the spectrum is well
represented by the single-phase model. This is also consistent with
the results of nearby non-XD clusters.  For the region (2), because
the radiative cooling time scale of the G1 emission is shorter than
the age of the Universe (See \S\ref{subsec:xrayproperties}), the gas
might be in a two-phase state. However, from the two-phase model
fitting, we found that the upper limit of the cool-phase flux is less
than 1\% of the hot-phase and the hot-phase temperature, 4.2 (3.5 --
5.3) keV, is again consistent with that obtained for the single-phase
model. Thus we consider that it is unlikely that the G1 gas is in a
two-phase state and that the virial temperature is much higher than
that derived from the single-phase model.

\cite{Czoske_etal_2001, Czoske_etal_2002} measured the redshift
distribution of galaxies in the direction of CL0024+17 and revealed
the presence of foreground and background groups of galaxies that
align along the line of sight.  \cite{Czoske_etal_2002} suggested from
their results that there was a high speed ($\sim 3000$ km/s) collision
between the CL0024+17 cluster and a second cluster with a mass about
half that of the main cluster $\sim3$ Gyr ago.  They suggested that
taking into account the projection effect of the second cluster mass
may solve the mass discrepancy problem.  They also mentioned that
during the collision the X-ray gas would be highly disturbed due to
propagation of shock waves \citep[e.g.][]{TAKIZAWA.1999} and
hydrostatic equilibrium of the gas component will break down, however,
after several Gyrs the gas will settle down to an equilibrium state.
\cite{Roettinger_etal_1996} noted that, based on numerical
simulations, the temperature structure is one of the strongest
indicator of recent merger activity, however, we do not find any
substantial temperature structure in the observational data as shown
in Figure \ref{fig:tprof}(a). We also found from the spectral analysis
of the G1 peak that the temperature is still higher than 2.6 keV at
the central $r<4\arcsec$ region. Thus our results show that the
cluster is in the state of several Gyr after the merging, and the gas
should have settled down to an equilibrium state and then trace the
underlying dark matter potential, which is supportive of
\cite{Czoske_etal_2002}'s scenario.

As discussed above, the total cluster mass is considered to be properly 
evaluated from the X-ray observations at the virial radius based on the 
consistency of the gas-mass fraction with the Universal baryon fraction. 
On the other hand, the large mass discrepancy detected  in the central region 
may indicate that there was a merging along the line of sight 
which disturbs the mass distribution in the cluster core region, as 
suggested by \cite{Czoske_etal_2002}. The G1 peak that we found in 
the X-ray image may be related with the merging.  In addition if we are 
seeing the two collided cores superposed along the line of sight, the 
mean redshift values may be slightly different between the cores.  
However we could not constrain the redshifts of the G1 and the 
surrounding region from the X-ray spectral data under the current 
limited photon statistics and the energy resolution for the Fe line 
emission. We expect that the future {\it ASTRO-E2} mission will 
constrain the line-of-sight structure of the cluster mass distribution 
with its finest spectral resolution.

Although the previous precise lens modelings \citep{Tyson_etal_1998,
Broadhurst_etal_2000} were performed with an a priori assumption that
CL0024+17 is an isolated single cluster of galaxies, our results have
provided new evidences which show that this is not the case.  One of
the strongest pieces of evidence is the consistency of the gas-mass
fraction of this cluster obtained from our X-ray results with the
universal baryon fraction.  In the case that the previous lens models
\citep{Tyson_etal_1998, Broadhurst_etal_2000} describe the cluster
mass distribution up to the virial radius correctly, the gas mass
fraction falls to $1/3$ of the universal baryon fraction.  Thereby it
will be a key subject to construct a new lens model based on the
current X-ray results and the results of the optical redshift survey
by \cite{Czoske_etal_2002} which suggests the existence of the merger
in the line of sight, and to examine the details of the physical
nature of the G1 component by further X-ray observations.

\section{Summary}
With {\it Chandra} we have performed the spectral and the spatial
analysis of the lensing cluster CL0024+17. The temperature profile is
consistent with being isothermal and the average temperature is $\sim
4.5$ keV. We detected a strong redshifted iron line with a
corresponding iron abundance of $\sim 0.8$ solar, which is one the
largest values among known clusters. We found that the radial X-ray
surface brightness distribution is not fitted with a single-component
$\beta$-model or an NFW-SSM model and that there is significant excess
emission centered at central bright elliptical galaxy.  The X-ray
surface brightness distribution was well described by a 2-dimensional
$\beta$-model of two extended components: the G1 component with a
small ($\sim 50$ kpc) core radius and the surrounding main cluster
component with a core radius $\sim 210$ kpc and whose center is
shifted by $\sim 80$ kpc from the G1 center. We derived the X-ray mass
within the arc radius assuming hydrostatic equilibrium and compared it
to the strong lensing mass. Our result clearly showed that the X-ray
mass is significantly smaller by a factor of 2--3. We then discussed
the possible cause of the mass discrepancy concerning the two-phase
spectral model, the cluster merging and the effect of the secondary
potential in the lensing effect. Although we do not rule out the
possibility that G1 is a remnant of a cluster merger, because of the
absence of any substantial temperature structure indicative of a
recent merger the gas seems to have relaxed in the cluster
potential. It is also true that G1 plays an indispensable role to make
the surface mass density of the lens supercritical, however it is not
sufficient to reconcile the large mass discrepancy.  Considering the
fact that the X-ray measurement of the gas-mass fraction at the virial
radius is consistent with the universal baryon fraction, we suggest
that it is important to further study the physical nature of the
cluster core and clarify its relation to the merging process, and
incorporate it into constructing a new lens model together with the
latest optical information of the galaxy distribution.

\acknowledgments We are grateful to S. Sasaki, T. Kitayama, and
K. Masai for useful discussions and to K. Mihara and K. Shimasaku for
suggestions related to the chemical evolution of galaxies. We thank
G. Soucail for providing us the CFHT image of the cluster. We also
thank P. G. Edwards for his careful review of the manuscript. N.O. is
supported by a Research Fellowship for Young Scientists from the
JSPS. N.O. acknowledges Hayashi Memorial Foundation for Female Natural
Scientists. We thank the anonymous referee for constructive comments.


\clearpage

\begin{figure}
\caption{X-ray image of CL0024+17 in the $0.5-7$ keV band obtained
with the {\it Chandra} ACIS-S3 with overlaid contours.  The image is
adaptively smoothed and corrected with the exposure map. The
background is not subtracted.  The image and the contours are
logarithmically scaled.  The point sources detected with the ROSAT
HRI, S1--S5 \citep{Soucail_etal_2000} are shown with the ROSAT error
box of $16\arcsec\times16\arcsec$. The boundary of the ACIS-S3 CCD
chip is shown with $8\arcmin\times8\arcmin$ box.}\label{fig:image}
\end{figure}


\begin{figure}
\epsscale{0.5}
\plotone{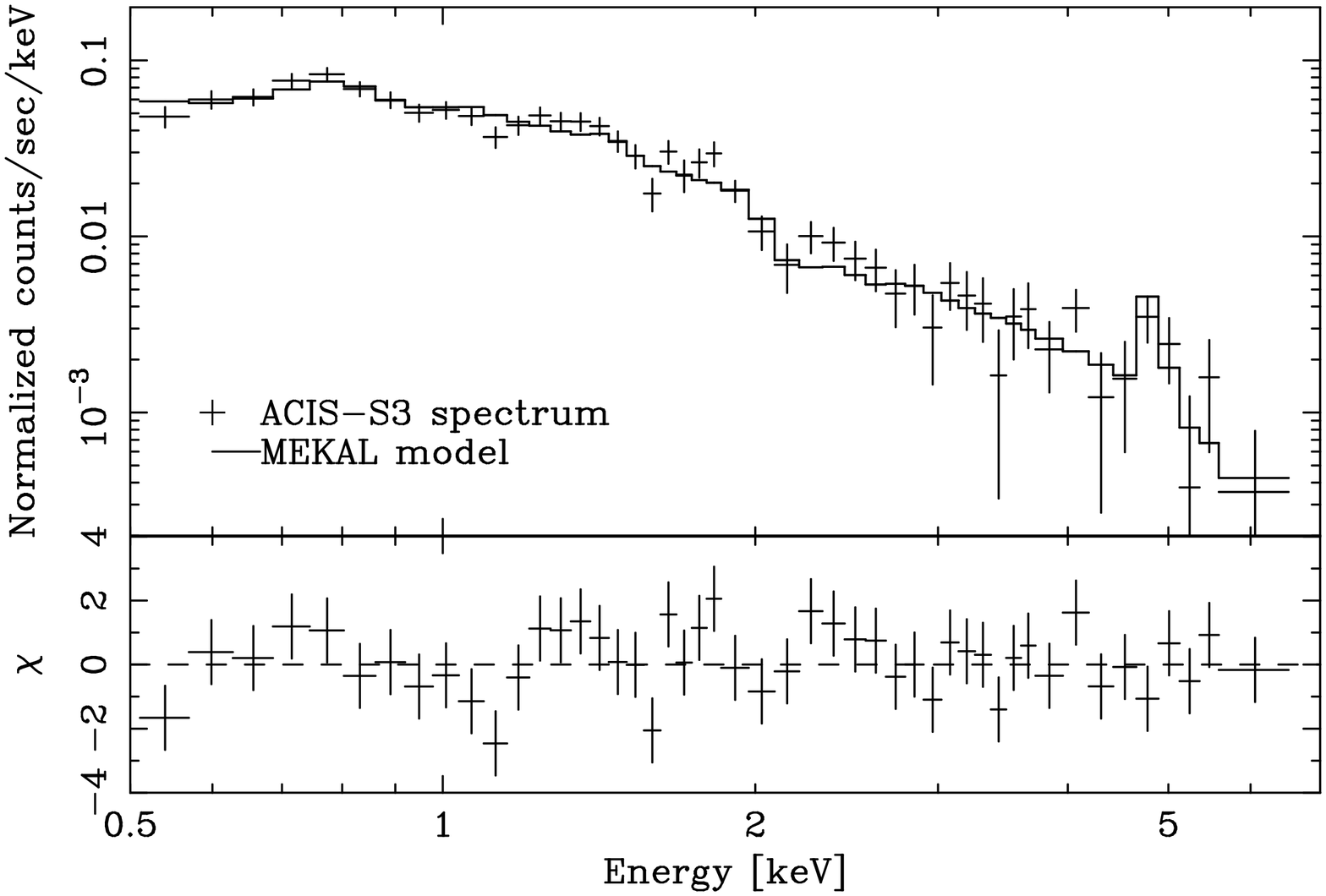}
\caption{{\it Chandra} ACIS-S3 spectrum of CL0024+17 ($r<1\arcmin.5$)
fitted with the MEKAL model. In the upper panel, the crosses denote
the observed spectrum and the step function shows the best-fit model
function convolved with the telescope and the detector response
functions. In the lower panel, the residuals of the fit in units of
$\sigma$ are shown.}
\label{fig:spectrum}
\end{figure}

\begin{figure}
\epsscale{1.0}
\plottwo{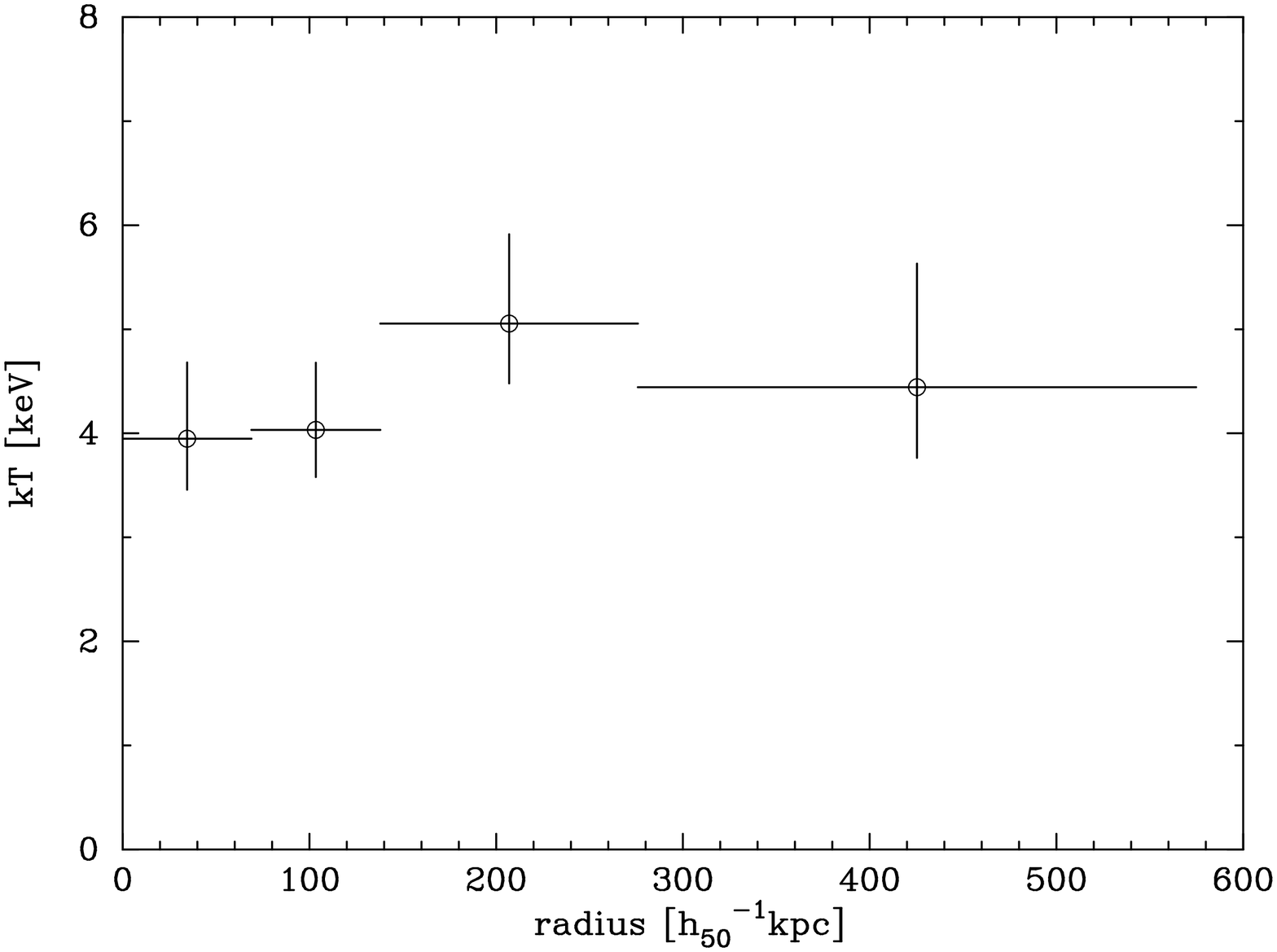}{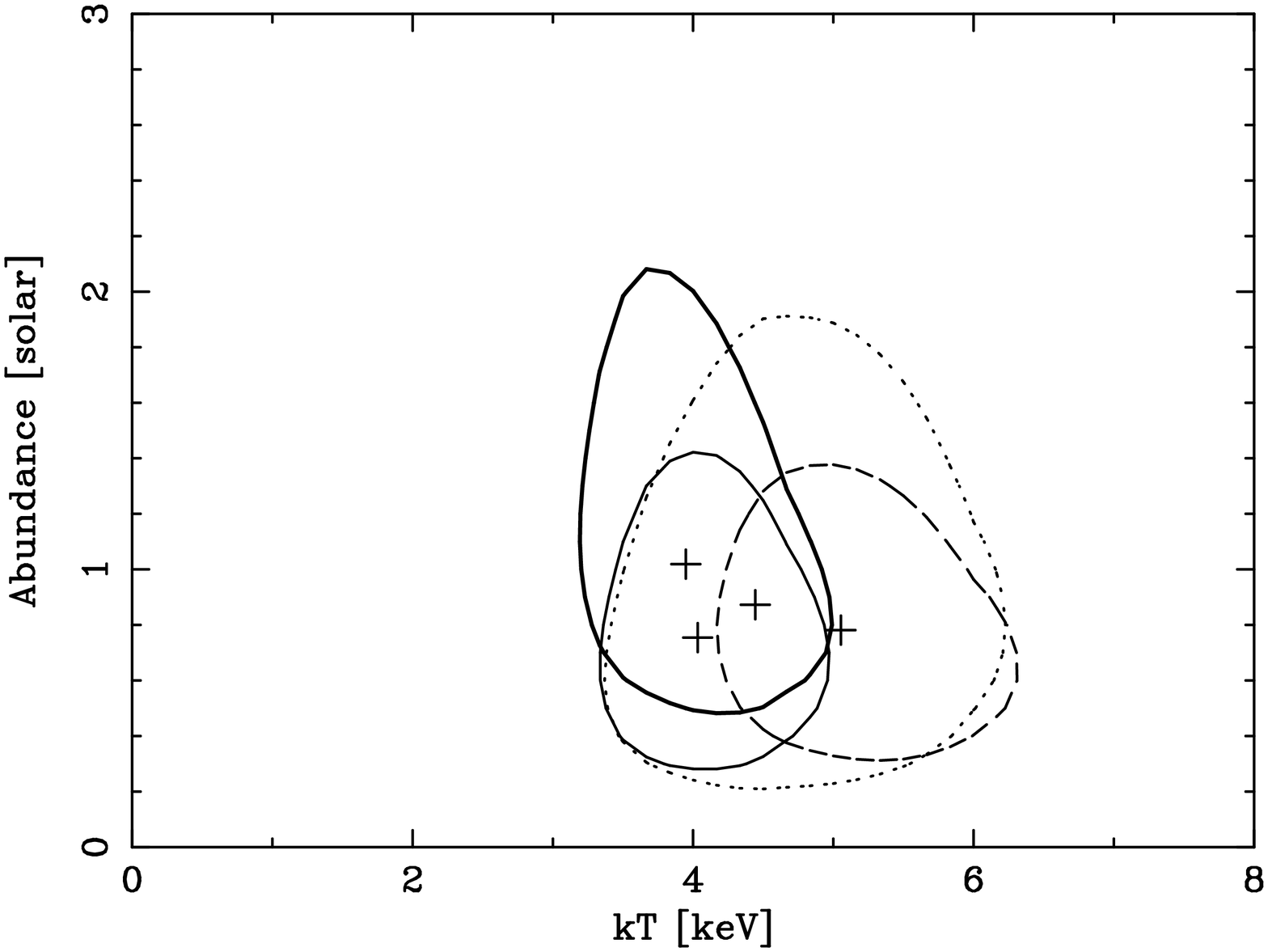}
\caption{(a) Radial temperature profile determined from the MEKAL
model fitting to the four annular regions. The vertical error bars are
$1\sigma$. (b) Confidence contours of the temperature and the metal
abundance for the same four annular regions used in the left
panel. The single-parameter 68\% error domains are shown as contours
with the thick-solid, thin-solid, thin-dashed, thin-dotted lines from
the inner to the outer annuli. }\label{fig:tprof}
\end{figure}

\begin{figure}
\plottwo{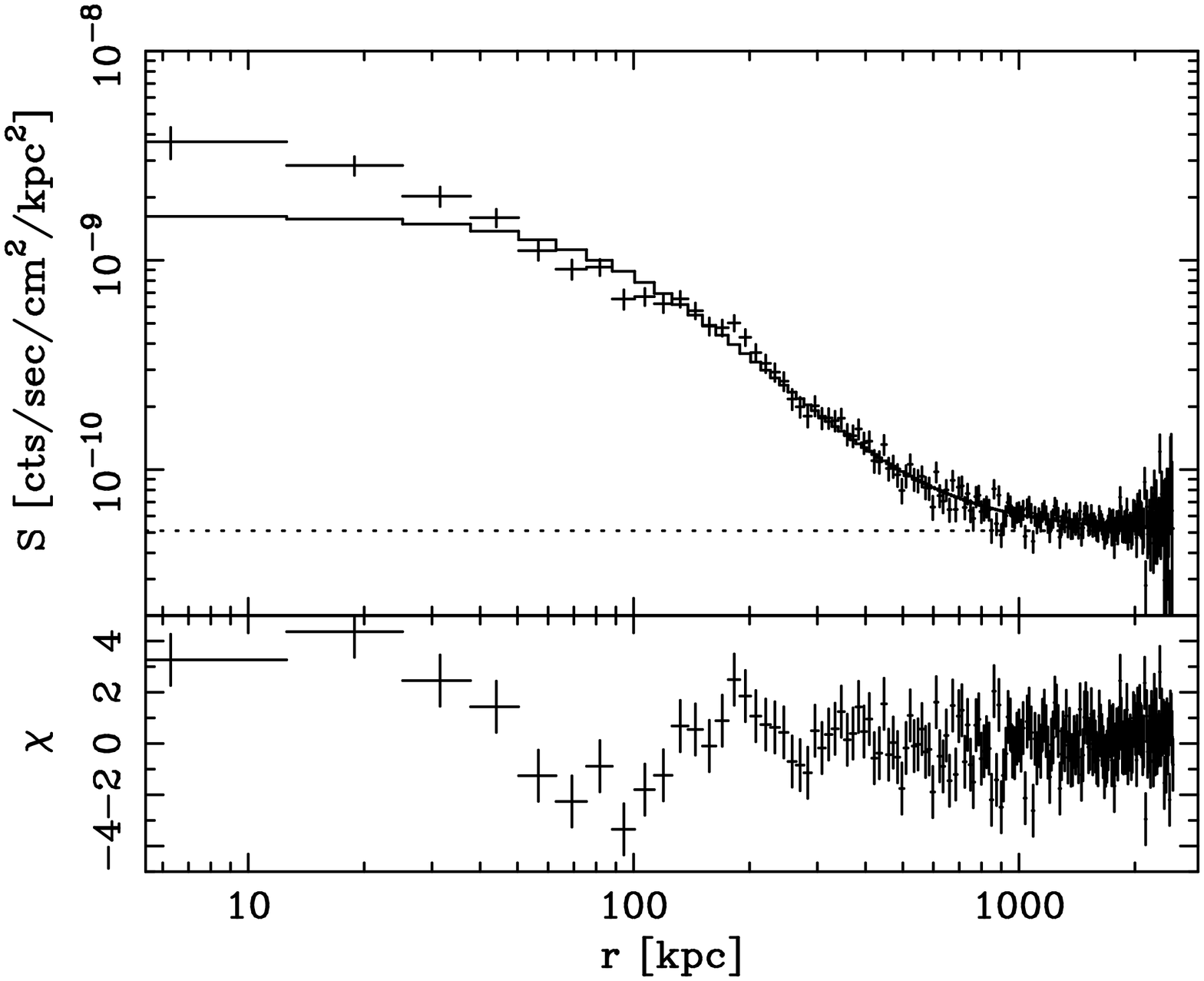}{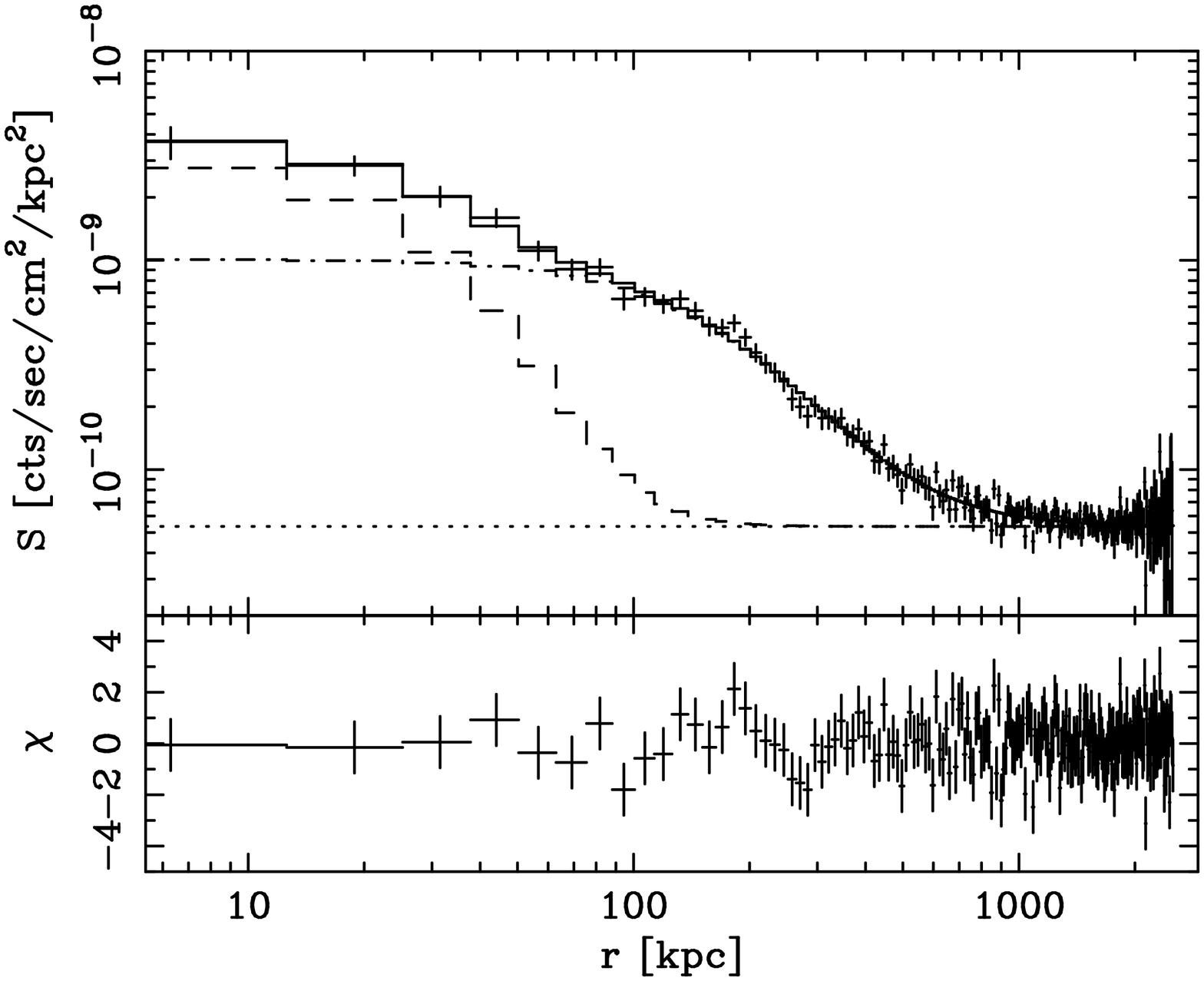}
\caption{Results of the 1-dimensional surface brightness distribution
fitting with the single $\beta$ model (a) and the double $\beta$ model
(b). In each panel, the crosses show the observed surface brightness
and the solid line shows the best-fit model. The constant background 
is shown with the horizontal dotted line.  For the double $\beta$
model, the inner and the outer components are respectively shown with
the dashed and the dash-dotted lines. }\label{fig:1dfit}
\end{figure}

\begin{figure}
\caption{Results of the 2-dimensional surface brightness distribution
fitting with the three $\beta$ profiles.  (a) {\it Chandra} ACIS-S3
image of the central $3\arcmin.3\times3\arcmin.3$ region of CL0024+17
in the $0.5-5$ keV energy range.  The small circles denote the
positions of the galaxies with $0.38 <z< 0.41$
\citep{Czoske_etal_2001}. A closer view of the central
$30\arcsec\times 30\arcsec$ region is also shown in the upper-left
panel. (b) The best-fit 2D image of the three $\beta$-models, overlaid
by the contours with logarithmic spacing.  (c) Residuals of the 2D
image fitting. (d) Residuals projected to the $x$- and $y$- directions
in units of $\sigma$. }\label{fig:2dfit}
\end{figure}

\begin{figure}
\epsscale{0.5}
\plotone{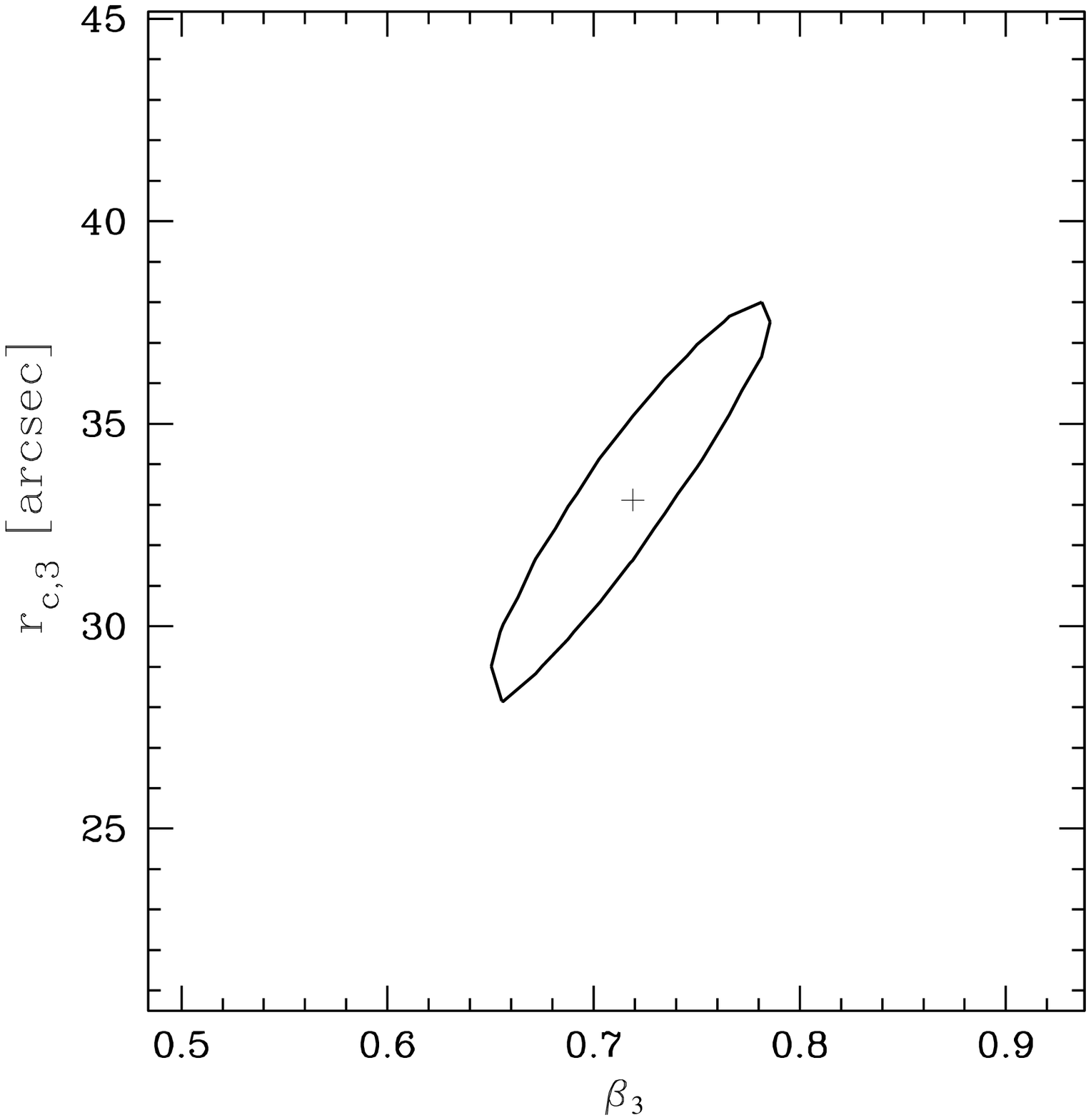}
\caption{$\chi^2$ contours of the 2-dimensional image fitting.
The 90\% single-parameter error domain for $\beta_3$ and $r_{c,3}$
of the cluster component is shown.
The position of the $\chi^2$ minimum is denoted with a cross.
}\label{fig:contour}
\end{figure}

\begin{figure}
\caption{X-ray mass density profiles for the $\beta$-model (the thick
line) and the NFW model (the thin line) of the main cluster component.
The 90\% error ranges are shown with the dashed and the dotted lines,
respectively. The arc radius of $220h_{50}^{-1}$kpc is denoted with
the arrow. The projected dark matter density profile derived by
\cite{Tyson_etal_1998} was plotted with the solid curve using the
fitted parameters given by \cite{Shapiro_Iliev_2000}. The critical
surface density, $\Sigma_{\rm crit}$, is indicated with the horizontal
dash-dot line. }\label{fig:densityprof}
\end{figure}

\begin{figure}
\caption{Critical lines for a source at $z=1.675$ are overlaid on the
B band image of CL0024+17 obtained with CFHT (the original image was 
described in \cite{Czoske_etal_2001}). North is up and east is
left. The multiple arc images are marked with A--E in the same manner
as \cite{Colley_etal_1996}. No critical line appears in the case of a
cluster lens for the main cluster potential (see
\S\ref{subsec:xraymass}). On the other hand, when adding the G1
potential the critical lines appear (the inner and the outer curves
are the radial and the tangential critical lines respectively),
however, it is not enough to reach to the southeast image (see
\S\ref{subsec:G1mass}). The object numbers of the central elliptical
galaxies in the catalog of \cite{Czoske_etal_2001} are also
shown. }\label{fig:critical_line}
\end{figure}

\clearpage
\begin{table}
\begin{center}
\caption{Results of the MEKAL model fitting to the overall spectrum}
\label{tab:spec}
\begin{tabular}{ll}\\\hline\hline
Parameter  & Value (90\% error) \\\hline
$N_{\rm H}$ (${\rm cm^{-2}}$) & $4.2\times10^{20}$ (F) \\
$kT$ (keV)  &   4.47 (3.93 -- 5.30)  \\
Abundance ($Z_{\sun}$) & 0.76 (0.45 --1.13) \\
Redshift & 0.395 (F) \\
Normalization\tablenotemark{a} & 5.72 (5.20 -- 6.25)  \\
$\chi^2/dof $ &47.5/45 \\
$f_{{\rm X}, 0.5-7}$ (${\rm erg/s/cm^2}$)\tablenotemark{b} & $4.1\times10^{-13}$ \\
$L_{{\rm X}, 0.5-7}$ (${\rm erg/s}$)     \tablenotemark{c}& $3.4\times 10^{44}$ \\
$L_{{\rm X}, {\rm bol}}$ (${\rm erg/s}$) \tablenotemark{d}& $5.1\times 10^{44}$ \\\hline
\end{tabular}
\end{center}
\tablenotetext{a}{Normalization factor for the MEKAL model, $\int n_e
n_{\rm H} dV/4\pi D_A^2 (1+z)^2$ in $10^{-18}{\rm cm^{-5}}$, 
where $D_A$ is angular size distance to the cluster.}  
\tablenotetext{b}{X-ray flux in the 0.5--7 keV band.}
\tablenotetext{c}{X-ray luminosity in the 0.5--7 keV band.}
\tablenotetext{d}{Bolometric luminosity.}
\tablenotetext{}{(F) Fixed parameters.} 

\end{table}

\begin{table}
\begin{center}
\caption{Results of the 1-D image fitting}
\label{tab:1dfit}
\begin{tabular}{llllll}\\\hline\hline
Model & (component) & $S_{0}\tablenotemark{a}$ & $\beta$ & $r_{c}$ & $\chi^2/dof$ \\
        & & ${\rm cts/s/cm^2/(h_{50}^{-1}kpc)^2}$ & & $\arcsec/h_{50}^{-1}$ kpc &  \\\hline
Single-$\beta$ & & $1.6^{+0.3}_{-0.2}\times10^{-9}$ & $0.55^{+0.04}_{-0.04}$ & $17.2^{+1.7}_{-1.6}/109^{+22}_{-20}$ & 
242.5/196 \\
Double-$\beta$ & (inner) & $2.8^{+0.8}_{-0.7}\times10^{-9}$  &1.0(F) & $7.0^{+1.4}_{-1.4}/45^{+10}_{-9}$ & 183.2/194 
\\
 & (outer) & $9.6^{+0.2}_{-0.1}\times10^{-10}$ &$0.66^{+0.08}_{-0.06}$ & $29.5^{+6.0}_{-5.0}/187^{+38}_{-31}$& 
\\\hline\hline
Model & $\alpha$  & $S_{0}\tablenotemark{a}$ & $B$ & $r_{s}$ & $\chi^2/dof$ \\
      & & ${\rm cts/s/cm^2/(h_{50}^{-1}kpc)^2}$  & & $\arcsec/h_{50}^{-1}$ kpc &  \\\hline
NFW-SSM & 1(F)    & $1.8^{+0.3}_{-0.2}\times10^{-9}$ & $8.3^{+0.8}_{-0.6}$& $65^{+17}_{-13}/415^{+107}_{-86}$ & 
229.3/196\\
\hline           
\end{tabular}
\end{center}
\tablenotetext{}{(F) Fixed parameters.} 
\tablenotetext{a}{Central surface brightness of the $\beta$-profile or the NFW-SSM model in the $0.5-5$ keV.}
\end{table}

\begin{table}
{\footnotesize 
\begin{center}
\caption{Results of the 2-D image fitting with the three $\beta$-models}
\label{tab:2dfit}
\begin{tabular}{lllllll}\\\hline\hline
Model & $i$ & Center position & $S_{0,i}\tablenotemark{a}$ & $\beta_i$ & $r_{c,i}$ & $L_{\rm X,bol}$ \\
component& &RA,Dec in J2000 &${\rm cts/s/cm^2/(h_{50}^{-1}kpc)^2}$  & & $\arcsec/h_{50}^{-1}$ kpc & erg/s \\\hline
G1 	& 1 & 00:26:36.0,+17:09:45.9(F)& $3.1^{+0.9}_{-0.6}\times10^{-9}$ & 1 (F) & $8.3^{+1.7}_{-1.4}/52^{+11}_{-9}$ 
& $5.5\times 10^{43}$\\
G2 	& 2 & 00:26:35.1,+17:09:38.0 (F)& $4.7^{+2.9}_{-2.5}\times10^{-9}$ & 1 (F) & 1.6/10 (F) & $3\times 10^{42}$\\
Cluster & 3 & 00:26:35.6,+17:09:35.2 \tablenotemark{b}& $8.7^{+1.2}_{-0.9}\times10^{-10}$ & $0.71^{+0.07}_{-0.06}$& 
$32.8^{+5.1}_{-4.7}/210^{+33}_{-30}$& $4.5\times 10^{44}$\\\hline
\end{tabular}
\end{center}
\tablenotetext{}{(F) Fixed parameters.} 
\tablenotetext{a}{Central surface brightness of the $\beta$-profile in the $0.5-5$  keV.}
\tablenotetext{b}{The 90\% errors are 
$\pm1\arcsec.3$ for RA and $\pm1\arcsec.5$ for Dec.}
}
\end{table}

\end{document}